\documentclass[%
 aip,
 amsmath,amssymb,
 reprint,%
 floatfix,%
]{revtex4-1}

\usepackage{graphicx}
\usepackage{dcolumn}
\usepackage{bm}
\usepackage{color}
\usepackage{comment}

\newcommand{\Ttrain}{T_{\mathrm{train}}}
\newcommand{\Ttest}{T_{\mathrm{test}}}
\newcommand{\ntrials}{n_{\mathrm{trials}}}
\newcommand{\nens}{N_{\mathrm{ens}}}

\definecolor{deep_blue}{RGB}{31, 119, 180}
\definecolor{deep_orange}{RGB}{255, 127, 14}
\definecolor{deep_green}{RGB}{44, 160, 44}
\definecolor{deep_red}{RGB}{214, 39, 40}
\definecolor{deep_purple}{RGB}{148, 103, 189}
\definecolor{deep_brown}{RGB}{140, 86, 75}
\definecolor{deep_pink}{RGB}{227, 119, 194}

\usepackage[colorlinks,linktoc=all]{hyperref}
\hypersetup{
  linkcolor=deep_pink,    
  citecolor=deep_green,   
  urlcolor=deep_red       
}
\usepackage[capitalise]{cleveref}

\newcommand{\x}{\mathbf{x}}

\newcommand{\rs}{\mathbf{r}}
\newcommand{\us}{\mathbf{u}}

\newcommand{\bs}{\mathbf{b}}

\newcommand{\A}{\mathbf{A}}
\newcommand{\Win}{\mathbf{W}_{\mathrm{in}}}
\newcommand{\Wout}{\mathbf{W}_{\mathrm{out}}}

\begin{document}






\title{Understanding the superiority of multi-model ensemble forecasts through reservoir computing}

\author{Daniel Estevez Moya}
\email{destevez@pks.mpg.de}
\affiliation{Max Planck Institute for the Physics of Complex Systems, Nöthnitzer Str. 38, 01187 Dresden, Germany}
\author{Francesco Martinuzzi}
\email{martinuzzi@pks.mpg.de}
\affiliation{Max Planck Institute for the Physics of Complex Systems, Nöthnitzer Str. 38, 01187 Dresden, Germany}
\author{Edmilson Roque dos Santos}
\email{edmilson.roque.usp@gmail.com}
\affiliation{Max Planck Institute for the Physics of Complex Systems, Nöthnitzer Str. 38, 01187 Dresden, Germany}
\author{Erick Alejandro Madrigal Solis}
\affiliation{Max Planck Institute for the Physics of Complex Systems, Nöthnitzer Str. 38, 01187 Dresden, Germany}%
\affiliation{Department of Physics, University of Technology, Dresden, Germany}
\author{Ernesto Estevez Rams}
\affiliation{Facultad de Física, Universidad de La Habana, San Lazaro y L. CP 10400, La Habana, Cuba}
\author{Holger Kantz}%
\affiliation{Max Planck Institute for the Physics of Complex Systems, Nöthnitzer Str. 38, 01187 Dresden, Germany}%

\date{\today}

\begin{abstract} 
    Weather forecasting and climate projection frequently use multi-model ensembles (MMEs) to improve short-term forecasts by averaging across models. However, this practice is often not well justified or validated. Using reservoir computing (RC) as a computationally efficient alternative to large-scale physical models, we assess the validity of the MME approach for chaotic time series. By training multiple randomly constructed RCs on the same dataset, we create a multi-model ensemble in which each model has its own unique error. These model errors lead to very different forecasting performances, with forecast error distributions that exhibit heavy tails. The arithmetic mean across forecasts from multiple models for the same target is usually closer to the ground truth than most individual forecasts, and further improvement is achieved by weighted arithmetic means where the weights are constructed based on each model's test-set performance. We show that iterated forecasts over many time steps deviate from the ground truth along the unstable manifold of the target point, in both directions, so that, if forecast errors were independent and had zero mean, the arithmetic mean forecast should approach the true target like $1/\sqrt{\nens}$ where $\nens$ is the size of the multi-model ensemble. We observe deviations from this behavior, which we attribute to the tails of the error distribution of random RCs.
\end{abstract}


\maketitle

\begin{quotation} 
        While multi-model ensembles (MMEs) forecasts in atmospheric science tend to outperform individual models, the underlying reasons for this improvement remain elusive. In this work, we adopt reservoir computing (RC) to investigate the mechanisms underlying MME short-term forecast performance. By generating ensembles of RCs with identical hyperparameters but different random realizations, we show that the superiority of the MMEs forecast is grounded in the geometric cancellation of errors projecting onto the unstable manifold of the chaotic attractor. Moreover, we provide a detailed statistical analysis of forecast error distributions and ensemble-size dependence. Although our guiding example is the time series generated by iterations of the Hénon map, the MMEs' forecast performance is also tested in higher-dimensional systems. Specifically, we show that the improved forecast performance holds in the folded towel map and in the continuous-time Lorenz~96 system. Our results provide the first steps towards an explanation for the improved performance of MMEs using a dynamical systems perspective.
\end{quotation}

\section{\label{mmerc:sec:introduction} Introduction}

    Multi-model ensembles (MMEs) are commonly used in weather forecasting and climate projections \citep{tebaldi2007use, leutbecher2008ensemble}. Different weather agencies and climate research centers possess their own general circulation models for the atmospheric dynamics, which can produce considerable spread across individual forecasts \citep{palmer2005representing}. While MMEs are primarily used to assess forecast uncertainty \cite{rougier2007probabilistic, knutti2010challenges, jebeile2020multimodel}, there is much evidence that the multi-model arithmetic mean on average provides better forecasts than each of the individual models \cite{MME_positive_2016}, so that in the IPCC reports the multi-model mean is usually adopted as the best available forecast \cite{IPCC5}. Specifically, \citet{lambert2001cmip1} note that ``\ldots no one model is “best” for all variables. There is some evidence that the “mean model” result, obtained by averaging over the ensemble of models, provides an overall best comparison to observations for climatological mean fields.'' 
    
    A simple mathematical model supporting the superiority of the multi-model arithmetic mean forecast is the following: assume that the true future value is $s_t$, and the $i$-th individual model predicts the value $\hat s_t^{(i)}= s_t+\epsilon_i$, then the arithmetic mean MME forecast is
    \begin{equation}\label{eq:MME_arithmetic}
    \langle \hat s_t^{(i)}\rangle_{\mbox{\scriptsize MME}} = s_t +\langle \epsilon_i\rangle_{\rm MME}\;.
    \end{equation} 
    If the individual model errors were mutually independent with zero mean, then 
    \begin{equation}\label{eq:MME_error}
    \langle \epsilon_i\rangle \propto\sigma_t/\sqrt{N_{ens}}\;,
    \end{equation} 
    where $N_{ens}$ is the ensemble size and $\sigma_t$ is the standard deviation of the individual forecast errors at time step $t$.

    Nevertheless, there are also strong objections against simple arithmetic averaging and merging models of different skill \cite{Weigel_risk,Wei_comparative_study}. Weather and climate models differ in their formulation in many ways, e.g., which physical processes are resolved or parametrized in which way, so they are really different models in terms of the number of degrees of freedom and equations of motion. The complexity of those models hinders an understanding of the superiority of the multi-model ensemble.
    
    To investigate this phenomenon without the overhead of high-dimensional climate models, reservoir computing (RC) stands out as a viable alternative \cite{nadiga2021reservoir}. RC is a machine learning architecture that has recently attracted attention for its ability to forecast nonlinear dynamics \citep{pathak2018modelfree, vlachas2020backpropagation}. Its main advantage is that training only affects the weights of the readout matrix, not the network's internal parameters \citep{jaeger2002tutorial, lukoeviius2009reservoir}.  A typical RC system has a large number of internal parameters, such as which links between the reservoir nodes are selected to be non-zero and which weights to assign to them \citep{lukoeviius2012practical}. These choices are usually made randomly, so that after fixing some hyperparameters, there is still a huge realm of possible realizations. Experience shows that different individual realizations of these random choices can lead to quite diverse forecast performances \citep{pathak2017using, haluszczynski2019good, chen2023proper}. In other words, there is a large variability in performance with respect to these choices, with no clear guidelines  \citep{martinuzzi2025minimal}. What at first sight appears to be a severe drawback turns out to be very beneficial for our setting: by fixing the hyperparameters and picking different realizations of the random components, one can easily and quickly construct an MME by training all RCs on the same training data set.

    In this paper, we study in detail the use of MMEs of RCs for improved short-term iterated (closed-loop) forecasts. The ease of generating the multi-model ensemble allows us to compare the properties of different models  and gain insight into \emph{why} the multi-model mean outperforms most individual forecasts. We find that while the forecast error of the multi-model mean still increases at a rate governed by the largest Lyapunov exponent of the underlying chaotic dynamics, its prefactor is significantly smaller than that of most individual forecasts. This is due to a cancellation of error vectors pointing in opposite directions along the unstable manifold of the chaotic trajectory. In other words, our results indicate that the simple mathematical model in \cref{eq:MME_error} remains somewhat relevant when the geometry of the unstable manifold is taken into account. Moreover, we analyze the variability in performance across identical hyperparameters by calculating the root mean squared forecast error across large numbers of RCs, and we find heavy tails in the probability of poorly performing models. We also show that the MME forecast improves with ensemble size and that using weighted means further enhances forecast performance. 
    
    In \cref{sec:ESNs} we briefly review the architecture of RC computing, and in \cref{sec:multistep} we discuss closed-loop multi-step forecasts and what we expect about error growth with forecast time. In \cref{sec:multi_model_ensemble_RC} we then show how to build the multi-model ensemble and how to combine the individual forecasts into an ensemble-averaged forecast. \cref{sec:henon} uses the well-known chaotic H\'enon map to generate the input data and exemplifies the superiority of the multi-model mean forecast. In \cref{sec:why} we discuss why the ensemble mean forecast has a high chance of being better than the individual forecasts. We show that forecast error vectors align along the unstable manifold of a dynamical system more or less symmetrically around the ground truth, so that on average, they can partly cancel each other. In \cref{sec:ndep} we present results on the improvement of multi-model ensemble forecasts as a function of ensemble size. We show that these features of MME forecasts are not restricted to the simple H\'enon map but also occur in higher-dimensional, more complex chaotic systems in \cref{sec:highdim}. \cref{sec:discussion_conclusions} contains a conclusion and outlook.

\section{\label{mmerc:sec:problem_statement} Multi-step forecasts by iterated Reservoir Computing}

    \subsection{\label{sec:ESNs} Echo state networks}
    
        Here we briefly recall the architecture of reservoir computing by echo state networks and introduce some notation, closely following \cite{Jaeger2001}. The reservoir consists of $N$ nodes and a directed weighted (i.e., non-symmetric) sparse adjacency matrix $\A$, forming a directed network with $n_l$ nonzero links, where self-loops are allowed. We use $n_l=0.01N^2$, i.e., a link density of 1\%, and each element $a_{ij}$ of the adjacency matrix is an i.i.d. random variable uniformly distributed in the interval $[-1, 1]$. The internal reservoir state at discrete time $t$ is denoted as an $N$-dimensional variable $\rs(t)$. An $N \times M$ input matrix $\Win$ is described by i.i.d. elements drawn from a uniform distribution in the interval $[-\sigma, \sigma]$, where $\sigma$ is the input scaling hyper-parameter. The input matrix couples the $M$-dimensional input $\{\us(t)\}_{t}$, linearly into the internal states, so that their dynamical update rule reads:
        \begin{equation}\label{eq:internal_states_evolution}
            \rs(t + 1) = (1-\alpha)\rs(t)+\alpha\tanh \big(\A \rs(t) + \Win \us(t) + \bs \big),
        \end{equation}
        where $\alpha$ is the leakage rate, and $\tanh$ is applied component-wise. Other choices for the activation function than $\tanh$ have been made in the more recent literature. The fixed shift $\bs$ consists of an $N$-dimensional vector whose entries are i.i.d. random variables drawn uniformly in the interval $[-1, 1]$ \citep{Platt2022}. 
    
        In this paper, we consider the input data $\us(t)$ always given by a deterministic dynamical system. As pre-processing, we shift and rescale the $M$-dimensional input data to $[-1,1]^M$ for the numerical stability of all operations in our reservoirs. Moreover, we consider $\alpha = 1$.
    
        The network should have the `echo state property' \cite{Jaeger2001}: driven by a sequence of inputs, the internal states $\rs(t)$ become independent of their initial conditions $\rs(0)$ for sufficiently large $t$. Setting the spectral radius $\rho$ of $\A$ (the largest absolute value of all its complex eigenvalues) below $1$ is a common rule of thumb for this, but it is in general neither sufficient nor necessary~\cite{yildiz2012revisiting, manjunath2013echo}. Whether the property holds depends on the input driving and not on $\A$ alone. Based on results in \cite{transienttimes} we fix the washout time to $1000$ update steps for spectral radii $\rho<0.9$. Due to the continued driving by the time-dependent inputs, the internal state synchronizes with the input signal in the sense of generalized synchronization \cite{Grigoryeva_2021_chaos_diff_sync}.
    
        After this synchronization has been achieved, the actual training of the network can start: we want to forecast the future of the input $\us(t)$, hence we use an affine function $f: \mathbb{R}^{N} \to \mathbb{R}^{N + 1}$, $f(\rs)=(\rs,1)$ to map the internal states to a feature vector for the ridge regression during training (this function can also be a nonlinear function). The $M \times (N+1)$ dimensional read-out matrix $\Wout$ is obtained by linear regression of the target time series onto the reservoir output $\hat{\us}(t) = \Wout f(\rs(t))$ such that
        \begin{align*}
            \frac{1}{\Ttrain}\sum_{t = 0}^{\Ttrain-1} \|\us(t+1)- \Wout f(\rs(t))\|^2 + \beta \|\Wout\|_F^2,
        \end{align*}
        where $\|\cdot\|$ and $\|\cdot\|_F$ are the Euclidean norm in $\mathbb{R}^M$ and $\|\cdot\|_F$ is the Frobenius norm, respectively, is minimized on a training set of length $\Ttrain$ of input-target pairs (supervised learning). The hyper-parameter $\beta$ is the regularizer parameter of the ridge regression. Once these weights have been obtained, the RC system can be used for forecasting the target.
    
        It is important to note that this is not a simple input-output task, because the internal state depends on the past inputs. Therefore, when using the system for forecasting, one must ensure that the new input is the natural continuation of the input time series that led to the current internal state.

    \subsection{\label{sec:multistep} Iterated (closed loop) forecasts and error growth}
    
        In many applications, like weather forecasting, one is interested in forecasting considerably far into the future. In typical situations, one predicts from the last known system state a small time step into the future and uses the outcome of this forecast as input for the next step, thereby iterating the prediction scheme many time steps, sometimes called closed loop forecasts. Given the reservoir output $\hat{\us}(t+1)$ as the forecast of $\us(t+1)$, one can iterate using as input for the next step the output $\hat{\us}(t)$ of the previous step.

        Forecast performance can be quantified by different metrics. We will in particular focus on the root mean squared  error (RMSE) defined as
        \begin{align}\label{eq:rmse_trials}
            \mathrm{RMSE}(k) = \Big( \frac{1}{\ntrials} \sum_{i = 1}^{\ntrials} \|\us^{i}(k) - \hat{\us}^{i}(k)\|^2 \Big)^{\frac{1}{2}},     
        \end{align}
        where the average is performed over $\ntrials$ forecast trials, $\us^i(k)$ denotes the $i-$th trial (or initial condition of the original system), and $k$ is the number of time steps ahead made by the iteration scheme described in \cref{sec:ESNs}. On average, the RMSE grows with the number of iterations until the forecast loses its value because the errors are of the order of the dynamical range of the data. This time is often called the forecast horizon or the valid prediction time \cite{vlachas2020backpropagation} and is a useful quantifier for the performance of a forecast scheme. 

        The reason for error growth is twofold: the last known state of the system is usually only known with some uncertainty, called the initial condition error. Even with a perfect forecast scheme, this error will be propagated into the future and, if the system is chaotic, will grow exponentially in time. For our numerically generated data, this error is the round-off error of the data. In addition, there is some model error, i.e., the calculation of a future state is done with a slightly wrong model compared to reality.  After a first iteration step, one can re-interpret the model error (leading to a slightly wrong image point) as an initial condition error for the next forecast step, i.e., model errors will be amplified by chaos or turbulence, but also iterating repeatedly with a wrong model will cause increasing differences between forecast and truth.

        As one expects for deterministic chaotic systems, we found that the RMSE grows roughly exponentially with the number of iterations into the future, where the exponent is close to the positive Lyapunov exponent $\lambda_1$ of the chaotic system. However, in order to find $\lambda_1$, one should perform a geometric average over forecast errors, while for the purpose of assessing forecast skill, an arithmetic average is better interpretable. However, the arithmetic mean can grow slightly faster than with $\lambda_1$ if the instability due to chaos is heterogeneous in the phase space. 

\section{\label{sec:multi_model_ensemble_RC} Multi-model ensemble reservoir computing}

    \begin{figure*}[ht]
        \centering
        \includegraphics[width=\textwidth]{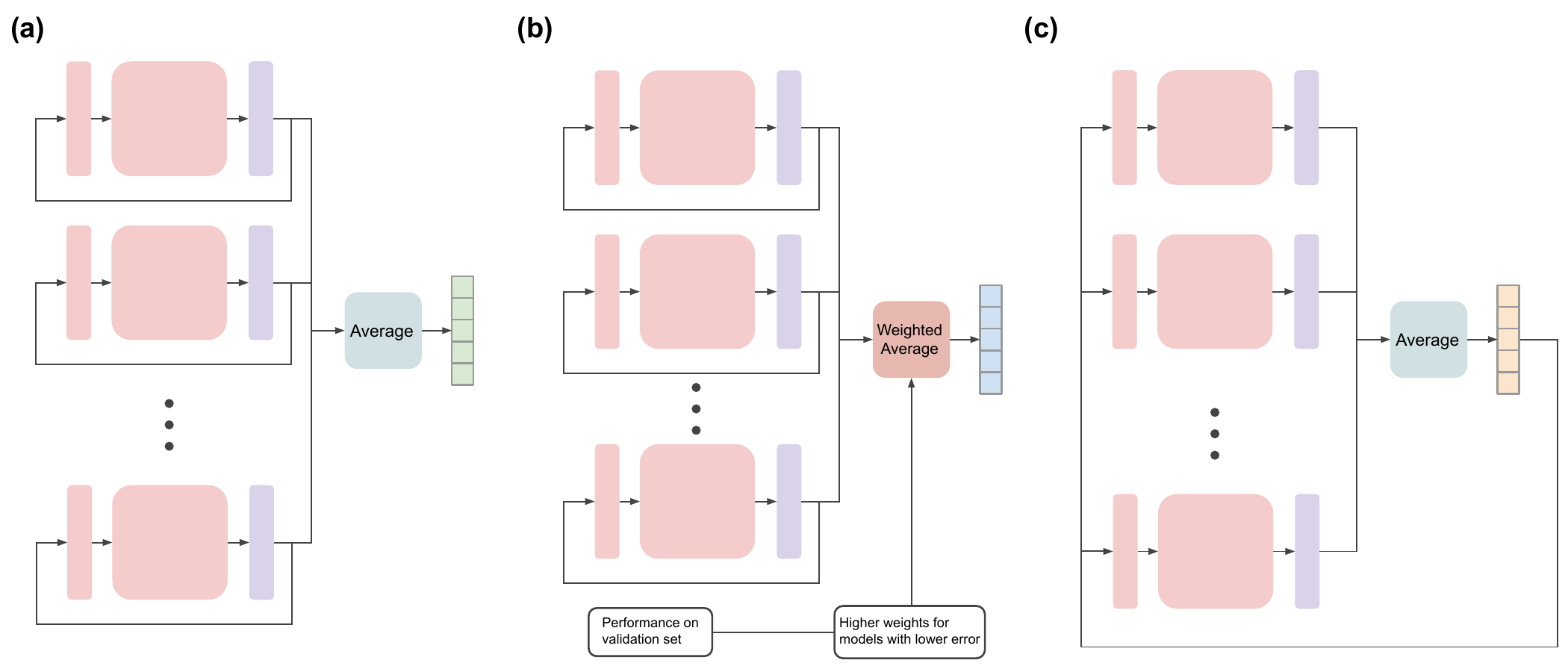}
        \caption{\textbf{Schematics for three multi-model ensemble strategies with reservoir computing.} a) The multi-model ensemble defined by the arithmetic average of the individual reservoirs' outputs. b) The weighted average ensemble is defined by a weighted average, where weights are higher for individual reservoirs with lower performance error in a validation set. In cases a) and b), each ensemble member runs independently and autonomously, as shown by the individual closed-loop schemes. c) Closed-loop ensemble autonomously utilizes the average of the individual reservoirs' output, i.e., each ensemble reservoir is fed by the average of the individual reservoirs' output of the previous step, by a closed-loop scheme as shown in the figure.}
        \label{mmerc:fig:mme}
    \end{figure*}

    The performance of RC computing depends sensitively on the hyperparameters $N$, $n_l$, $\rho$, $\sigma$, and $\alpha$, but even more on the random choices made for the adjacency matrix $\A$, the input weights $\Win$, and the bias term \citep{pathak2017using, chen2023proper}. We turn the latter, which looks like a disadvantage, into a benefit, because it offers a very convenient way to create a multi-model ensemble:  we construct and train $\nens$ RC models for the same input data with the same hyper-parameters, but different random choices of the adjacency matrix and the input weights, see \cref{mmerc:fig:mme} for an illustration. Like in multi-model ensembles in weather forecasting and climate science, one can now combine the individual ensemble member forecasts to a multi-model ensemble forecast as follows: Let us denote by the index $j$ the properties of the $j$th reservoir inside our ensemble of $\nens$ RCs, e.g., $\rs^{(j)}$, $\A^{(j)}$, $\Win^{(j)}$, $\bs^{(j)}$ and $\Wout^{(j)}$. Then we define three types of ensemble averages:

    \textbf{Arithmetic average (a):} We consider the arithmetic average of the individual ensemble member forecasts made for the same time instance $t$:
    \begin{align}\label{eq:arithmetic_mean_ensemble}
        \langle\hat{\us}\rangle^{a}(t) = \frac{1}{\nens} \sum_{j = 1}^{\nens} \hat{\us}^{(j)}(t),
    \end{align}
    this can be visualized in \cref{mmerc:fig:mme} a).
    
    \textbf{Weighted average (w):} Rather than a simple arithmetic average, we also consider a weighted average of the individual ensemble member forecasts:
    \begin{align}\label{eq:weighted_mean_ensemble}
        \langle\hat{\us}\rangle^{w}(t) = \frac{1}{\sum_{j = 1}^{\nens} \omega_j} \sum_{j = 1}^{\nens} \omega_j \hat{\us}^{(j)}(t),
    \end{align}
    where the weights $(\omega_j)_{j = 1}^{\nens}$ are defined during testing (validation). For our study, we have fixed $\omega_j = 1/\sqrt{RMSE_{\Ttest}^{(j)}(k=5)}$, i.e., the inverse square root of the 5-step ahead forecast error of the ensemble member $j$ over $\Ttest$ iterates during testing. This is an \textit{ad hoc} choice without any claim of optimality. The scheme is illustrated in \cref{mmerc:fig:mme} b). 
    
    A further, small improvement to the forecast is achieved if, in the iteration for a multi-step forecast, not the individual forecast for each RC is fed back, but instead the multi-model ensemble mean forecast. This means that at each step, we compute the ensemble mean of the forecasts from all RCs and feed it back to all RCs to compute the next step. In this way, the whole ensemble remains close together, while we still make use of the model variety at each step. Mathematically, we define
    
    \textbf{Closed-loop ensemble (CLE):} 
    \begin{eqnarray}\label{eq:closed_loop_ensemble}
       &\rs^{(j)}(t + 1) = \tanh{\big(\A^{(j)} \rs^{(j)}(t) + \Win^{(j)} \textcolor{deep_blue}{\langle \hat{\us}\rangle(t)} + \bs^{(j)} \big)} \nonumber\\
       &\hat{\us}^{(j)}(t + 1) = \Wout^{(j)} f(\rs^{(j)}(t + 1)), 
    \end{eqnarray}
    where the blue term $\textcolor{deep_blue}{\langle \cdot \rangle}$ corresponds to either the arithmetic mean $(a)$ or weighted mean $(w)$ defined in \cref{eq:arithmetic_mean_ensemble,eq:weighted_mean_ensemble}. This scheme is illustrated in \cref{mmerc:fig:mme} c). 
    
    In principle, if at one step the mean over the ensemble produced a really bad forecast, the whole ensemble could drift away from the true future. However, it seems that this is never the case, as the RMS forecast error of this scheme is still slightly better than taking the multi-model mean forecast, where each individual RC feeds back into itself. However, there is only a little benefit from this, since all models operate in close vicinity to the correct future state. One could therefore linearize every individual reservoir around this point of reference, and then the arithmetic mean of the future states of all RCs is essentially the average of the images of the arithmetic mean of all current states. We find that the prediction horizon is extended by 1-2 time steps. We therefore do not consider feeding back the ensemble mean as a major further improvement and will not present any numerical results in the following.

\section{Guiding example: H\'enon map}\label{sec:henon} 

    Let us consider the H\'enon map \cite{Henon}, formally a 2-dimensional chaotic map, which, however, can be easily rewritten in the following way:
    \begin{equation}\label{eq:henon}
      x(t + 1) = 1 - 1.4 x(t)^2 + 0.3 x(t-1).
    \end{equation}
    In this example, one sees the strength of the RC approach: while the knowledge of $x(t)$ alone is insufficient to predict $x(t+1)$, the reservoir `memorizes' the previous input $x(t-1)$ in its internal state, so that with the scalar driving by $x(t)$ one can achieve excellent forecasts for $x(t+1)$ and can iterate these many times before the forecast error becomes larger than 0.1 (for rescaled data with variance about 1). 
    
    \begin{figure}
    \includegraphics[width=1.0\linewidth]{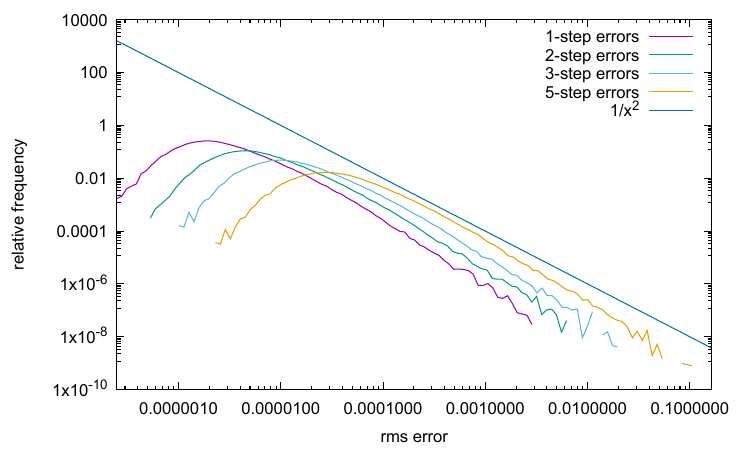}
    \caption{\label{fig:rmse-distributions}
    Distributions of RMS-errors of 157500 individual RCs trained on the very same data set of the H\'enon map. 
    Hyper-parameters: 100 nodes, 150 links, spectral radius 0.95, input scaling 1
    }
    \end{figure}
    
    \begin{figure}[ht]
    \includegraphics[width=1.0\linewidth]{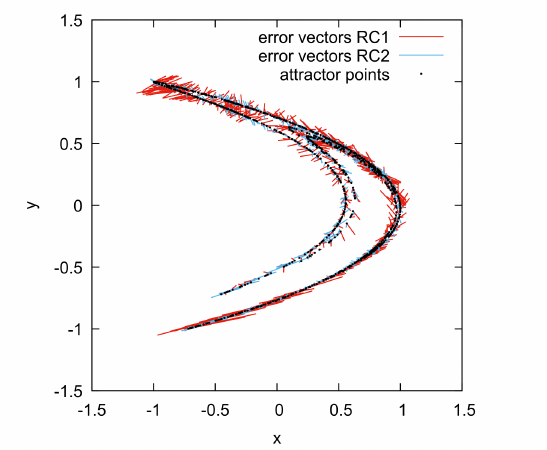}
    \caption{\label{fig:individual_errors} Individual 2-step forecast error vectors (magnified by a factor of 15) shown as red (blue) lines connecting the `true future' (black dot)  and the prediction made by the red (blue) RC model after training, on the very same data set. }
    \end{figure}
    
    Nonetheless, the performances of RCs, even with identical hyperparameters, can vary tremendously. In Fig.\ref{fig:rmse-distributions} we show distributions of forecast errors for iterated predictions. The tails towards large errors, i.e., bad performance, seem to be power law-like, so that rarely but with non-zero probability the random components of an RC can cause the model to be really bad. As expected, the distributions are shifted towards larger errors when the number $k$ of steps into the future increases.
    
    In \cref{fig:individual_errors} we show forecast errors as vectors connecting the truth and the predicted value for forecasts on the same data set from the H\'enon map, for 2 randomly chosen RCs. There are regions on the attractor where one sees essentially only blue vectors, i.e., in these parts of the phase space, the `blue' model has larger model errors, while in other (many more) regions the red error vectors are dominant. This exemplifies that, like in weather forecasting, even if the `blue' model might perform better on average, there are parts of the phase space where the `red' model is better.
      
    While the component of error vectors along the stable direction (transverse to the attractor) will be damped out in future iterations, the component into the unstable direction, tangent to the attractor, will exponentially grow. Additional model errors in the following iteration steps are small but will be amplified as well and accumulate.
    
\begin{figure}
        \includegraphics[width=1.0\linewidth]{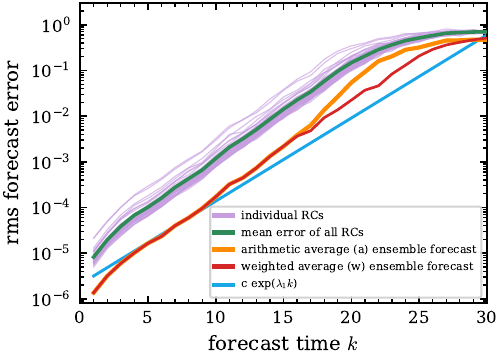}
        \caption{\label{fig:henon_multimodelwithfeedback} Root mean squared prediction errors of RCs averaged over 500 prediction trials for the H\'enon map: light purple curves show the RMSEs of each single one of 100 RCs, and the green curve their mean. Orange curve: RMS forecast error when using the arithmetic ensemble mean as forecast; red curve: using the weighted ensemble mean. Error growth is slightly faster than with the Lyapunov exponent of the H\'enon map, as shown by the blue line $c\,e^{\lambda_1 k}$.}
    \end{figure}

    We show the evolution of the RMSE with the number of time steps $k$ into the future for the H\'enon map in \cref{fig:henon_multimodelwithfeedback}. All errors grow roughly exponentially in $k$, with an exponent slightly larger than the maximal Lyapunov exponent. The errors of the individual RCs for each $k$ exhibit a large spread of almost 2 orders of magnitude; among the $\nens=100$ different RCs, some have really poor performance. Still, the RMSE of the arithmetic mean forecast \cref{eq:arithmetic_mean_ensemble} is almost as low as the best individual forecasts. There is further improvement when using the weighted average Eq.(\ref{eq:weighted_mean_ensemble}). In our numerical experiments with input data truncated after the 5th digit, we achieve prediction horizons in the range of 20 to 25 for individual RCs, and a prediction horizon of $\approx 25$ when using the multi-model ensemble mean as forecast, see \cref{fig:henon_multimodelwithfeedback}. 
    
\section{Why the multi-model mean is the better forecast}\label{sec:why}

\begin{figure}
        \includegraphics[width=1.0\linewidth]{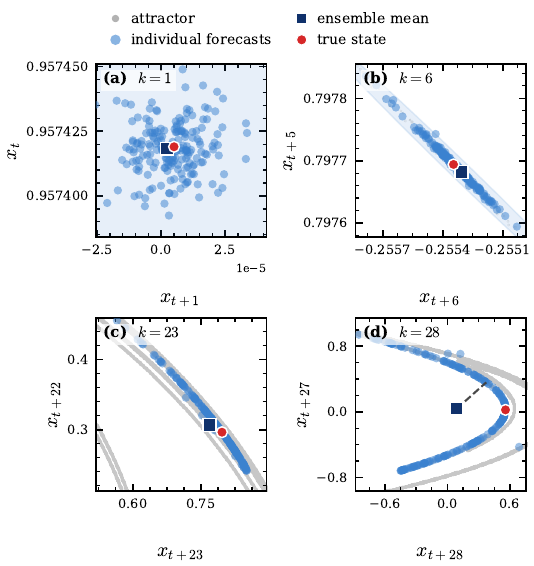}
        \caption{\label{fig:error-clouds}The individual forecasts of 200 randomly created RC models (blue circles) together with the true point (red disc) and the arithmetic mean over the members (dark blue square) for $k=1$, 6, 23, and 28 iteration steps of iterated forecasts. Each panel is zoomed to the forecast cloud, with the two axes scaled independently; notice that the scale changes with $k$, i.e., deviations of the individual forecasts from the truth become larger. While the 1-step model errors have no preferred direction in phase space, the forecasts distribute along the unstable manifold (the attractor, in gray) for higher iterates, until at $k=28$ the curvature of the manifold places the arithmetic mean outside it, in the concave gap off the attractor.}
    \end{figure}

    We analyze RC-forecast errors for the H\'enon map \cref{eq:henon} in more detail. This map has a strange attractor of fractal dimension $D_f\approx 1.34$ and one positive Lyapunov exponent (LE) $\lambda_1\approx 0.42$ and one negative LE of $\lambda_2\approx -1.62$. All values are known only through numerical simulations. In \cref{fig:error-clouds}, we show for a single prediction trial the forecast points of the different RCs together with the truth and with their arithmetic mean \cref{eq:arithmetic_mean_ensemble}, to understand in detail how the errors evolve over time. The phenomenon described above is nicely visible: The deviations of the forecasts from the truth are more or less isotropic in both space directions and very small in amplitude when $k=1$ is small. So the model errors without further iteration are in all space directions of comparable magnitude and somewhat random, depending on how each individual RC has learned the underlying dynamics. In this case, the simple mathematical model \cref{eq:MME_arithmetic} grounded on the i.i.d. of the model errors could be approximately applicable. The model errors have probability distributions whose density is supported on the plane.

    When the forecast horizon $k$ increases, they spread out (i.e., larger forecast errors) along the unstable manifold of the attractor in the neighborhood of the true future point, while perpendicular to that, their spread remains small. The relevant feature of this spread is that it goes along \textit{both directions} of the unstable manifold. Therefore, it is evident that the average of the $k$-step forecasts from different models is closer to the truth than most individual forecasts. Interestingly, this suggests that the simple mathematical model in \cref{eq:MME_arithmetic} needs to be adapted. The assumption of i.i.d. model errors is only approximately valid when the geometry of the unstable manifold is explicitly considered. This implies that a more robust mathematical model should track model errors via a conditional probability distribution supported along the unstable manifold. Note that the arithmetic mean, however, moves away from the true point perpendicular to the attractor when the spread  becomes so large that the curvature of the unstable manifold becomes relevant, see the bottom row of \cref{fig:error-clouds}. 

\begin{figure}
        \includegraphics[width=1.0\linewidth]{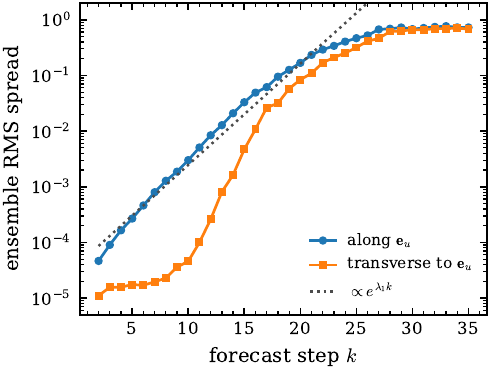}
        \caption{\label{fig:alignment} Root mean squared spread of the ensemble forecast errors, resolved along the local unstable direction $\mathbf{e}_u$ (blue) and transverse to it (orange), as a function of the forecast step $k$ for the H\'enon map, averaged over forecast trials. The along-manifold spread grows exponentially at essentially the largest Lyapunov rate ($\propto e^{\lambda_1 k}$, dotted), while the transverse spread stays far smaller, so that the cloud of forecast errors is confined to the one-dimensional unstable manifold. The anisotropy is largest (a factor of $\approx 65$) around $k=10$ and decreases for $k\gtrsim 17$, where the curvature of the manifold becomes relevant.}
    \end{figure}

    We can track this geometric picture quantitatively. At each forecast step $k$, every ensemble member produces a forecast that differs from the true point $\x(t+k)$ by an error vector. At that true point, we determine the local unstable direction $\mathbf{e}_u$, i.e., the direction along which the dynamics stretches small displacements. We take a tangent vector and evolve it with the Jacobian of the map along the trajectory that leads to $\x(t+k)$, which makes it line up with the most expanding direction. We then split each member's error vector into a component along $\mathbf{e}_u$ and a component transverse to it, and compute the root-mean-square size of each over the ensemble, averaged across many forecast trials. The result is shown in \cref{fig:alignment}. The spread along $\mathbf{e}_u$ grows exponentially at essentially the largest Lyapunov rate $\lambda_1$ (dotted line), while the transverse spread stays one to two orders of magnitude smaller, the two differing by up to a factor of $\approx 65$ around $k=10$. In other words, the cloud of forecast errors collapses onto the one-dimensional unstable manifold, exactly as the phase-space clouds in \cref{fig:error-clouds} suggest. Because the errors are moreover spread to both sides of the true point along this manifold, their arithmetic mean largely cancels. The collapse weakens only for $k\gtrsim 17$, where the spread grows large enough that the curvature of the manifold sets in and the transverse component starts to grow, the same regime in which the arithmetic mean begins to leave the attractor.

\section{Forecast performances as function of ensemble size $\nens$}\label{sec:ndep}

    Under the assumption that the model errors inside our ensemble of RCs are independent and have zero mean, the RMSE of the arithmetic mean forecast is expected to shrink like $1/\sqrt{\nens}$ as stated in Eq.(\ref{eq:MME_error}). This assumption is too rough to be true. It is well known that in typical dynamical systems, the local stretching rate can fluctuate considerably as a function of position on the attractor. This is the reason why finite-time Lyapunov exponents \cite{FTLE} fluctuate. In terms of forecast errors, this means that depending on where on the attractor the target of prediction is located, the forecasts of the very same RC will have different error magnitudes for identical initial condition errors. This property is inherited by the ensemble: In regions of strong local divergence, the ensemble of forecasts is expected to be more spread out than in regions of weak local divergence or even convergence. Actually, this is also known from weather forecasting: On the regional scale, there are weather situations where forecasts are very precise, and others where forecasts have much larger errors, as it is seen in the ensemble forecasts\cite{palmer2005representing}.\footnote{Notice that these ensemble forecasts are NOT multi model ensembles, but based on ensembles of initial conditions which are created by small perturbations of the reconstructed current state of the atmosphere}. So, while we still expect that forecast errors of independently constructed models are independent across single forecast trials, their variances may fluctuate from trial to trial. Therefore, in a multi-trial setting, the forecast errors are expected to be independent but not identically distributed. Still, this will not affect the average behavior as a function of $\nens$.
    
    Eq.(\ref{eq:MME_error}) relies on the additional assumption that the distribution of the independent errors $\epsilon_i$ has finite variance. In our way to construct the ensemble of RCs, the distribution of their forecast errors (already for a single trial) has long tails towards exceptionally big errors, Fig.\ref{fig:rmse-distributions}. This finding is consistent with error distributions shown in \cite{Griffith_2019_distribution_errors}. This fact has two consequences. First, the performance of a single ensemble depends much on whether or not such a badly performing model is a member of this ensemble. This can be compensated for by considering mean forecast errors, where we also average over many ensembles, especially when ensembles have only a few members. Secondly, the fact that individual errors can be so big, even if finite, seems to slow down the convergence of $\frac{1}{\nens}\sum \epsilon_i$.

\begin{figure}
        \includegraphics[width=1.0\linewidth]{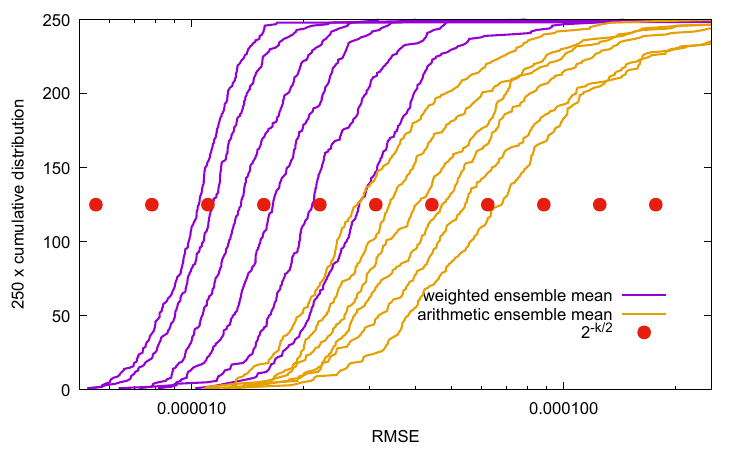}
        \caption{\label{fig:n_dependence} Cumulative distributions of root mean squared 5-step errors as functions of ensemble sizes, for the plain arithmetic mean forecasts (orange) and for the weighted arithmetic mean forecasts (purple) for ensembles of sizes $N_{ens}=10, 20, 40, 80, 160, 320$ from right to left, based on 250 different multi-model ensembles each. The red bullets show how the medians of the distributions should behave if the errors decreased as
        $1/\sqrt{\nens}$.}
    \end{figure}

    In Fig.~\ref {fig:n_dependence}, we show the cumulative distributions of 5-step forecast errors for ensemble forecasts as a function of ensemble size $\nens$ (H\'enon map). To be more specific, for every fixed ensemble size $\nens$, we create 250 such ensembles with randomly constructed RCs. For each realization of an ensemble, we calculate the RMSE of the arithmetic mean forecast Eq.(\ref{eq:arithmetic_mean_ensemble}) (orange) and of the weighted arithmetic mean forecast Eq.(\ref{eq:weighted_mean_ensemble}) (purple) 5 steps ahead. At the end, we perform a rank ordering of these mean forecast errors and plot rank versus error, which yields the empirical cumulative distributions of these errors. First, it is evident that the larger the ensemble size $\nens$, the more the distributions are shifted towards smaller errors. This supports our claim that a multi-model ensemble is better the larger the ensemble.
    
    Secondly, we see that the weighted arithmetic mean Eq.(\ref{eq:weighted_mean_ensemble}), where the weight factor of every individual model in the ensemble is taken as the inverse of the square root of the training set error of the model, is performing better than the plain arithmetic mean, simply because bad ensemble members contribute less to the forecast. Finally, suppressing these bad forecasts also narrows the error distributions, i.e., it reduces the ensembles' sensitivity to bad models. The median of these distributions can be read off as the error values at rank 125. The red circles located there symbolize a reduction of $1/\sqrt{2}$. If Eq.(\ref{eq:MME_error}) held, the distributions should be shifted by $1/\sqrt{2}$ when doubling the ensemble size. As one can see, the reduction of errors is weaker than this for the plain arithmetic mean, and is also weaker for the weighted arithmetic mean. While this failure of Eq.(\ref{eq:MME_error}) deserves deeper analysis, we think that it is related to the fat tail of the rms-error distributions of individual RCs. 
    
\section{Higher-dimensional systems}\label{sec:highdim}

    In this section, we show that multi-model ensembles also improve forecasts when the phase space dimension is larger than 2. 
    
    \subsection{Folded towel map}
    
        \begin{figure}
            \includegraphics[width=1.0\linewidth]{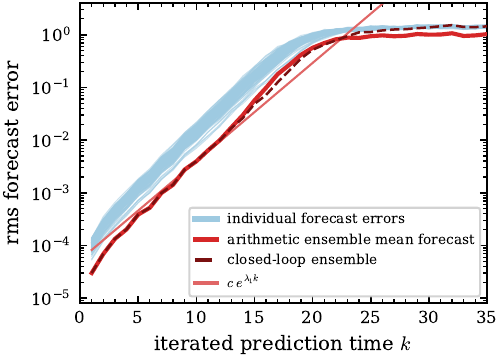}
            \caption{\label{fig:foldedtowel}As in \cref{fig:henon_multimodelwithfeedback}, but for the folded towel map \cref{eq:foldedtowel}. As inputs, we use the 3-dimensional state vectors of the map plus bilinear combinations of their components. As output, we predict the full state vector, and RMSEs are computed using the Euclidean distance in phase space. Ensemble size $N_{ens}=100$, RMSEs averaged over 500 prediction trials.}
        \end{figure}
        
        In Fig.\ref{fig:foldedtowel} we show forecast errors for the folded towel map\cite{foldedtowel} defined as:
        \begin{eqnarray}\label{eq:foldedtowel}
          x(t + 1)&=&3.6x(t)(1-x(t))-0.05(y(t)+0.35)(1-2z(t))\nonumber\\
          y(t + 1)&=&0.1\left((y(t)+0.35)(1-2z(t))\right)(1-1.9x(t))\nonumber\\
          z(t + 1)&=&3.78z(t)(1-z(t))+0.2y(t)
        \end{eqnarray}
        This system has 2 positive Lyapunov exponents of $\lambda_1=0.434$ and $\lambda_2=0.379$, so it is more complex than the H\'enon map. Its attractor has a fractal dimension of 2.24 (Lyapunov or Kaplan-Yorke dimension). 
        When using only the $x$-component as inputs, we do not find a good performance of our RCs. We therefore use the full 3-dimensional state vector as inputs and train 3 different sets of output weights for a forecast of the full state vector at time $t+1$. Since the prediction of the $y$-component still has forecast errors which are about 100 times larger than the other two, we extend the input vector further and feed in also all bilinear terms of the three variables. We use these so that we can track the RMSEs on a similar range of $k$ values as in the H\'enon system. Again, we observe that the multi-model ensemble mean forecast performs better than almost all of the individual reservoirs, which contribute to the mean value (Fig.\ref{fig:foldedtowel}), which implies a cancellation of the individual errors. 


        Since the state vectors of this map are 3D, we cannot easily visualize the error clouds as in the H\'enon map. In order to verify that they are also attracted to the unstable manifold, we perform dimension estimates of the clouds of forecast vectors, using the Grassberger--Procaccia correlation dimension $D_2$ \cite{GrassbergerProcacciaD2}. Let us consider a single prediction trial. While the forecasts of the different RCs after one step ($k=1$) fill a 3-dimensional volume around the true target point, the dimension of the corresponding forecast vectors after several steps, $k > 5$, shows a crossover to $D_2=2$. Because of the 2 positive Lyapunov exponents, this is exactly the dimension of the unstable manifold. So, for the H\'enon map, we therefore conjecture that the error vectors connecting the ground truth and the different RCs' forecasts are inside the unstable manifold as $k$ increases.

\subsection{Lorenz 96 system}

\begin{figure*}[ht]
        \centering
        \includegraphics[width=0.9\textwidth]{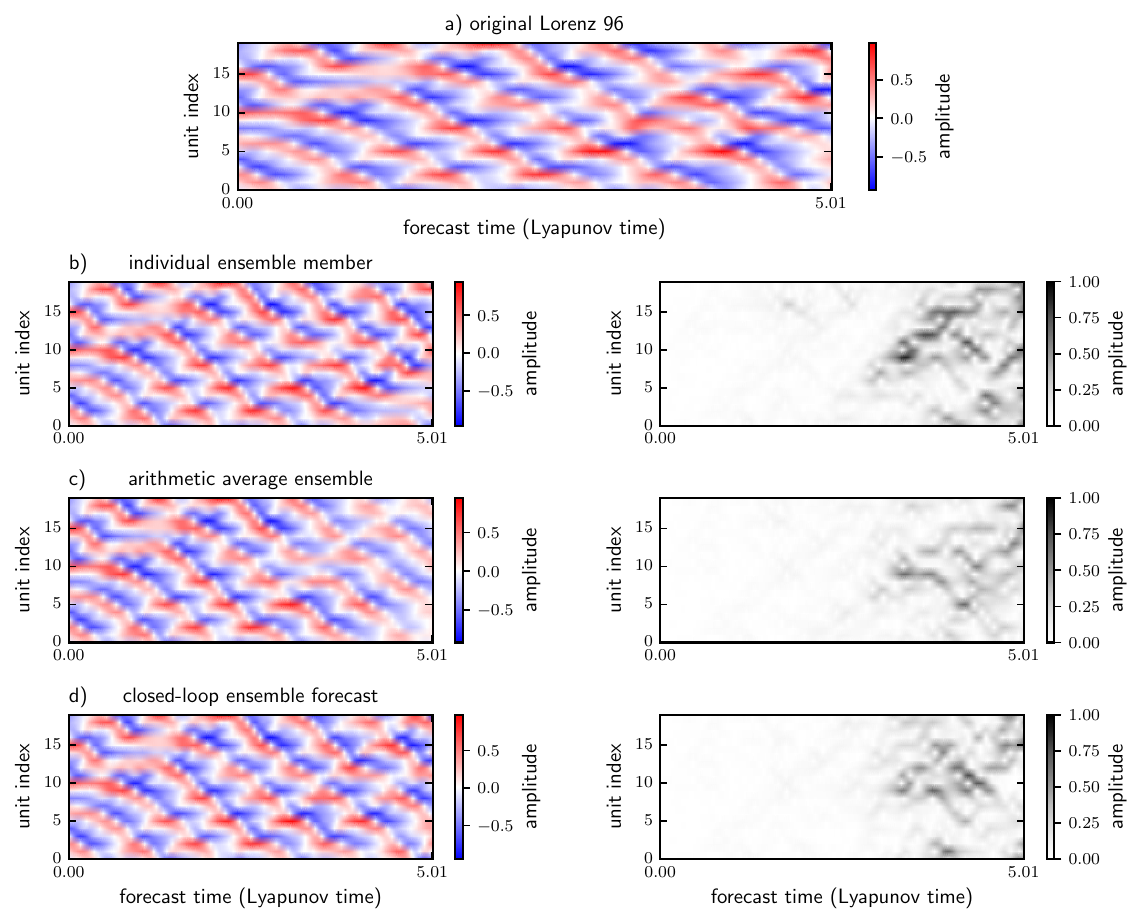}
        \caption{\textbf{Multi-model ensembles provide short-term forecasting of Lorenz 96 chaotic spatiotemporal trajectory.} a) Colormap of the original spatiotemporal trajectory, where color represents the state of each unit in the interval $[-1, 1]$. The remaining panels are divided into columns: the left panels show the colormap of the spatiotemporal forecast trajectory of each model, while the right panels show the colormap of the pointwise absolute error between the forecast and the original trajectory, shown in a). b) An individual ensemble member (chosen for visualization purposes). c) Arithmetic average ensemble forecast. d) Closed-loop ensemble forecast (also using arithmetic average in Eq.(\ref{eq:closed_loop_ensemble})). The hyper-parameters are $N = 5000$, $n_l = 0.01$, $\rho = 0.5$, $\sigma = 0.1$, and $\beta = 10^{-2}$.}
        \label{fig:spatiotemporal}
    \end{figure*}

    Following our initial motivation in weather forecasting, in this section, we show the MME performance for the short-term forecasting of spatiotemporal chaos. Instead of iterating over a discrete map, we consider the equations of motion of the Lorenz 96 system \cite{lorenz1996predictability} used to model the large-scale behavior of the mid-latitude atmosphere:
    \begin{align}\label{eq:Lorenz96}
        \dot{x}_j = (x_{j + 1} - x_{j - 2}) x_{j - 1} - x_j + F, \quad j = 0, \dots, M - 1,
    \end{align}
    where $F$ is the forcing term, and periodic boundary conditions $x_{-1} = x_{M - 1}$, $x_{-2} = x_{M - 2}$. For our purposes, we consider $F = 5$ and $M = 20$ as used in \cite{Chiara_ref_lorenz96_params}, which has a maximum Lyapunov Exponent of $\lambda_1 \approx 0.54$. We solve \cref{eq:Lorenz96} from random initial conditions uniformly distributed in the interval $[-1, 1]^{M}$ using a 4th-order Runge-Kutta numerical scheme and time step $h = 0.01$. The system is integrated over $1005 \tau_{\lambda_1}$ time, after transient effects were discarded, $T_{trans} = 100 \tau_{\lambda_1}$, where $\tau_{\lambda_1} = 1/{\lambda_1}$ corresponds to the Lyapunov time. The training data consists of the first $\Ttrain = 1000 \tau_{\lambda_1}$ time period, whereas the testing is the remaining part. The $M$-dimensional input data at time $t$ is given by $\us(t):= (x_0(t h), \dots, x_{M -1}(t h)) \in \mathbb{R}^{M}$ from the sampled time series, which is normalized to be in the set $[-1, 1]^{M}$. For this particular example, the nonlinear function $f(\rs)$ follows the choice of \cite{vlachas2020backpropagation} such that
    \begin{align*}
    f_i(\rs) = \begin{cases}
        \rs_i, \quad \mathrm{i = 1, \dots, N/2}, \\
        \rs_i^2, \quad \mathrm{otherwise}.
    \end{cases}   
    \end{align*}
    Moreover, the hyperparameters were selected via grid search to minimize short-term forecasting error across 10 different initial conditions. Specifically, the optimal hyperparameters are selected by minimizing the maximum RMSE over ten initial conditions and ten ensemble members, and are listed in the caption of \cref{fig:spatiotemporal}. \cref{fig:spatiotemporal} shows that MMEs provide a successful short-term forecasting of the chaotic spatiotemporal trajectory for at least 2.5 Lyapunov times, before the pointwise error starts to be noticeable. 
    
\begin{figure}
        \includegraphics[width=1.0\linewidth]{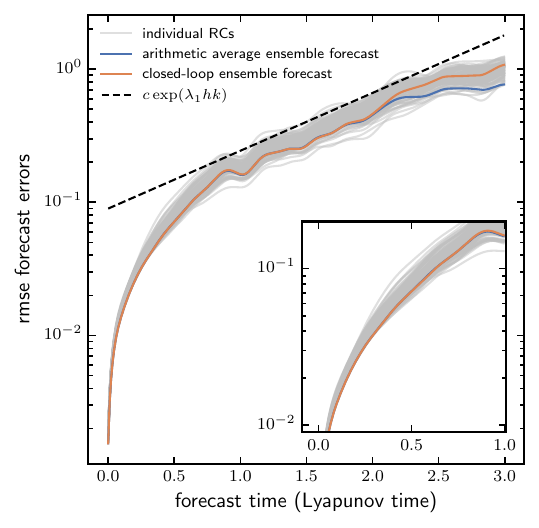}
        \caption{\label{fig:rmse_lorenz96} Root mean square error versus forecast time (quantified in Lyapunov time $\tau_{\lambda_1}$) of 100 ensemble members (in gray), arithmetic average (in blue), and closed-loop ensemble (in orange) with respect to 10 prediction trials. The black dashed line corresponds to the linear growth associated with the maximum Lyapunov Exponent. The multi-model ensembles' performance errors grow more slowly than most of the individual forecasts for most of the 3 Lyapunov time periods. As confirmed by the inset panel, which shows the zoomed-in region up to one-Lyapunov time. The hyper-parameters are those used in \cref{fig:spatiotemporal}}
    \end{figure}

    \cref{fig:spatiotemporal} is included for illustrative purposes. A quantitative analysis of the performance of the multi-model ensembles in terms of RMSE with respect to 10 prediction trials is shown in \cref{fig:rmse_lorenz96}. The MMEs provide a successful short-term forecasting, where the error grows more slowly compared to the individual ensemble members, which is confirmed by the inset panel in \cref{fig:rmse_lorenz96}. Interestingly, differing from the discrete maps case, the error between the reservoir trajectories and the reference grows in two different regimes. At first, the error growth is much faster than the exponential growth associated with the maximum Lyapunov exponent, persisting for one Lyapunov time, due to the high dimensionality of the system. Then, it decreases and follows the expected growth, as illustrated by the black dashed line. 
    
\section{\label{sec:discussion_conclusions} Discussion and conclusions}

    The use of MMEs through RC has been introduced in the literature before; for instance, H. Jaeger and collaborators have exploited the idea to take the mean of reservoir output for classification problems \cite{Jaeger_mme_mention_2007}. However, the approach has not yet gained widespread popularity. Only more recently, a growing body of work suggests that ensemble strategies constitute a promising direction for improving the robustness and predictive performance of RC across a variety of applications. This ranges from investigations, motivated by weather forecasting applications \cite{nadiga2021reservoir}, to modeling the evolution of dynamic aperture for hadron machines \cite{Casanova_2023_hadron_application}. Moreover, with the need for more energy-efficient hardware through physical realization of RC, ensemble mean over identical reservoir units is relevant to reduce the impact of different sources of noise affecting a physical reservoir system, in particular, dynamical and observational noise \cite{nakamura2026ensemblereservoircomputingphysical}.
    
    This work demonstrated that RC can reveal interesting geometric and statistical features that explain why an MME achieves good short-term forecasting performance, while mimicking different physical models used in weather forecasting and climate projection. Our numerical results confirm that MMEs generally outperform the majority of their individual ensemble members. Our contributions are  two-fold. First, we identify that the mathematical model in \cref{eq:MME_arithmetic} is a good proxy of the reason for better MME's performance, once the geometry of the unstable manifold of the original dynamics is taken into account. Error vectors of different models are not independent, but seem to be independently distributed inside the unstable manifold so that they can partly cancel each other. Second, as expected, increasing the ensemble size improves overall short-term forecasting performance, confirming that MME is better the larger the ensemble. The performance can be further improved by using weighted means instead of the arithmetic mean. An interesting research direction is to formalize those numerical findings theoretically.

\section{Authors Contributions}
\textbf{Daniel Estevez Moya:} Investigation (equal); Methodology (equal); Software (equal); Validation (equal); Writing – review \& editing (equal); 
\textbf{Francesco Martinuzzi:} Investigation (equal); Methodology (equal); Validation (equal); Writing – review \& editing (equal); \textbf{Edmilson Roque dos Santos:} Investigation (equal); Methodology (equal); Software (equal); Validation (equal); Visualization (equal); Writing – review \& editing (equal);
\textbf{Erick A. Madrigal Solis:} Validation; Reviewing;
\textbf{Ernesto Estevaz Rams:} Conceptualization, Validation; Writing - review \& editing;
\textbf{Holger Kantz:} Conceptualization; Investigation; Methodology; Validation; Visualization; 
Software; Writing - original draft preparation;
\section{Data availability}

    The data that support the findings of this study are available from the corresponding author upon reasonable request.

\section{Acknowledgments}

    The authors thank Thomas G. de Jong and Francesco Sorrentino for enlightening discussions. E.R.S. acknowledges support by FAPESP Grant No. 23/13706-0.

\bibliographystyle{aipnum4-1}
\bibliography{references_chaos}

@book{IPCC5,
  author    = {{IPCC}},
  title     = {{Climate Change 2013: The Physical Science Basis}},
  subtitle  = {Contribution of Working Group I to the Fifth Assessment Report of the Intergovernmental Panel on Climate Change},
  editor    = {Stocker, T. F. and Qin, D. and Plattner, G.-K. and Tignor, M. and Allen, S. K. and Boschung, J. and Nauels, A. and Xia, Y. and Bex, V. and Midgley, P. M.},
  publisher = {Cambridge University Press},
  address   = {Cambridge, United Kingdom and New York, NY, USA},
  year      = {2013},
  pages     = {1535}
}

@article{jebeile2020multimodel,
  title = {Multi-model ensembles in climate science: Mathematical structures and expert judgements},
  volume = {83},
  ISSN = {0039-3681},
  url = {http://dx.doi.org/10.1016/j.shpsa.2020.03.001},
  DOI = {10.1016/j.shpsa.2020.03.001},
  journal = {Studies in History and Philosophy of Science Part A},
  publisher = {Elsevier BV},
  author = {Jebeile,  Julie and Crucifix,  Michel},
  year = {2020},
  month = oct,
  pages = {44–52}
}

@article{knutti2010challenges,
  title = {Challenges in Combining Projections from Multiple Climate Models},
  volume = {23},
  ISSN = {0894-8755},
  url = {http://dx.doi.org/10.1175/2009JCLI3361.1},
  DOI = {10.1175/2009jcli3361.1},
  number = {10},
  journal = {Journal of Climate},
  publisher = {American Meteorological Society},
  author = {Knutti,  Reto and Furrer,  Reinhard and Tebaldi,  Claudia and Cermak,  Jan and Meehl,  Gerald A.},
  year = {2010},
  month = may,
  pages = {2739–2758}
}

@article{rougier2007probabilistic,
  title = {Probabilistic Inference for Future Climate Using an Ensemble of Climate Model Evaluations},
  volume = {81},
  ISSN = {1573-1480},
  url = {http://dx.doi.org/10.1007/s10584-006-9156-9},
  DOI = {10.1007/s10584-006-9156-9},
  number = {3–4},
  journal = {Climatic Change},
  publisher = {Springer Science and Business Media LLC},
  author = {Rougier,  Jonathan},
  year = {2007},
  month = jan,
  pages = {247–264}
}

@techreport{Jaeger2001,
  author      = {Jaeger, H.},
  title       = {The `echo state' approach to analysing and training recurrent neural networks},
  institution = {GMD -- German National Research Institute for Computer Science},
  number      = {GMD Report 148},
  year        = {2001}
}

@article{Henon,
  author  = {H{\'e}non, M.},
  title   = {A two-dimensional mapping with a strange attractor},
  journal = {Communications in Mathematical Physics},
  volume  = {50},
  pages   = {69--77},
  year    = {1976}
}

@article{foldedtowel,
  author  = {R{\"o}ssler, O. E.},
  title   = {An equation for hyperchaos},
  journal = {Physics Letters A},
  volume  = {71},
  pages   = {155--157},
  year    = {1979}
}

@article{pathak2018modelfree,
  title = {Model-Free Prediction of Large Spatiotemporally Chaotic Systems from Data: A Reservoir Computing Approach},
  volume = {120},
  ISSN = {1079-7114},
  url = {http://dx.doi.org/10.1103/PhysRevLett.120.024102},
  DOI = {10.1103/physrevlett.120.024102},
  number = {2},
  journal = {Physical Review Letters},
  publisher = {American Physical Society (APS)},
  author = {Pathak,  Jaideep and Hunt,  Brian and Girvan,  Michelle and Lu,  Zhixin and Ott,  Edward},
  year = {2018},
  month = jan 
}

@misc{nakamura2026ensemblereservoircomputingphysical,
      title={Ensemble Reservoir Computing for Physical Systems}, 
      author={Yuma Nakamura and Tomoyuki Kubota and Yusuke Imai and Sumito Tsunegi and Hirofumi Notsu and Kohei Nakajima},
      year={2026},
      eprint={2601.21807},
      archivePrefix={arXiv},
      primaryClass={math.DS},
      url={https://arxiv.org/abs/2601.21807}, 
}

@article{Grigoryeva_2021_chaos_diff_sync,
  title = {Chaos on compact manifolds: Differentiable synchronizations beyond the Takens theorem},
  author = {Grigoryeva, Lyudmila and Hart, Allen and Ortega, Juan-Pablo},
  journal = {Phys. Rev. E},
  volume = {103},
  issue = {6},
  pages = {062204},
  numpages = {12},
  year = {2021},
  month = {Jun},
  publisher = {American Physical Society},
  doi = {10.1103/PhysRevE.103.062204},
  url = {https://link.aps.org/doi/10.1103/PhysRevE.103.062204}
}

@article{vlachas2020backpropagation,
title = {{Backpropagation algorithms and Reservoir Computing in Recurrent Neural Networks for the forecasting of complex spatiotemporal dynamics}},
journal = {Neural Networks},
volume = {126},
pages = {191-217},
year = {2020},
issn = {0893-6080},
doi = {https://doi.org/10.1016/j.neunet.2020.02.016},
author = {P.R. Vlachas and J. Pathak and B.R. Hunt and T.P. Sapsis and M. Girvan and E. Ott and P. Koumoutsakos}
}

@article{nadiga2021reservoir,
author = {Nadiga, Balasubramanya T.},
title = {Reservoir Computing as a Tool for Climate Predictability Studies},
journal = {Journal of Advances in Modeling Earth Systems},
volume = {13},
number = {4},
pages = {e2020MS002290},
doi = {https://doi.org/10.1029/2020MS002290},
url = {https://agupubs.onlinelibrary.wiley.com/doi/abs/10.1029/2020MS002290},
note = {e2020MS002290 2020MS002290},
year = {2021}
}

@article{lambert2001cmip1,
  title = {{CMIP1 evaluation and intercomparison of coupled climate models}},
  volume = {17},
  ISSN = {0930-7575},
  url = {http://dx.doi.org/10.1007/PL00013736},
  DOI = {10.1007/pl00013736},
  number = {2–3},
  journal = {Climate Dynamics},
  publisher = {Springer Science and Business Media LLC},
  author = {Lambert,  S. J. and Boer,  G. J.},
  year = {2001},
  month = jan,
  pages = {83–106}
}

@inbook{lukoeviius2012practical,
  title = {A Practical Guide to Applying Echo State Networks},
  ISBN = {9783642352898},
  ISSN = {1611-3349},
  url = {http://dx.doi.org/10.1007/978-3-642-35289-8_36},
  DOI = {10.1007/978-3-642-35289-8_36},
  booktitle = {Neural Networks: Tricks of the Trade},
  publisher = {Springer Berlin Heidelberg},
  author = {Lukoševičius,  Mantas},
  year = {2012},
  pages = {659–686}
}

@article{pathak2017using,
  title = {Using machine learning to replicate chaotic attractors and calculate Lyapunov exponents from data},
  volume = {27},
  ISSN = {1089-7682},
  url = {http://dx.doi.org/10.1063/1.5010300},
  DOI = {10.1063/1.5010300},
  number = {12},
  journal = {Chaos: An Interdisciplinary Journal of Nonlinear Science},
  publisher = {AIP Publishing},
  author = {Pathak,  Jaideep and Lu,  Zhixin and Hunt,  Brian R. and Girvan,  Michelle and Ott,  Edward},
  year = {2017},
  month = dec 
}

@article{chen2023proper,
  title = {Proper choice of hyperparameters in reservoir computing of chaotic maps},
  volume = {56},
  ISSN = {1751-8121},
  url = {http://dx.doi.org/10.1088/1751-8121/acfb54},
  DOI = {10.1088/1751-8121/acfb54},
  number = {41},
  journal = {Journal of Physics A: Mathematical and Theoretical},
  publisher = {IOP Publishing},
  author = {Chen,  Wei and Gao,  Jian and Yan,  Zixiang and Xiao,  Jinghua},
  year = {2023},
  month = sep,
  pages = {415702}
}

@article{haluszczynski2019good,
  title = {Good and bad predictions: Assessing and improving the replication of chaotic attractors by means of reservoir computing},
  volume = {29},
  ISSN = {1089-7682},
  url = {http://dx.doi.org/10.1063/1.5118725},
  DOI = {10.1063/1.5118725},
  number = {10},
  journal = {Chaos: An Interdisciplinary Journal of Nonlinear Science},
  publisher = {AIP Publishing},
  author = {Haluszczynski,  Alexander and R\"{a}th,  Christoph},
  year = {2019},
  month = oct 
}

@misc{jaeger2002tutorial,
  title={{Tutorial on training recurrent neural networks, covering BPPT, RTRL, EKF and the echo state network approach}},
  author={Jaeger, Herbert},
  volume={5},
  number={1},
  year={2002},
  publisher={GMD-Forschungszentrum Informationstechnik Bonn}
}

@article{lukoeviius2009reservoir,
  title = {Reservoir computing approaches to recurrent neural network training},
  volume = {3},
  ISSN = {1574-0137},
  url = {http://dx.doi.org/10.1016/j.cosrev.2009.03.005},
  DOI = {10.1016/j.cosrev.2009.03.005},
  number = {3},
  journal = {Computer Science Review},
  publisher = {Elsevier BV},
  author = {Lukoševičius,  Mantas and Jaeger,  Herbert},
  year = {2009},
  month = aug,
  pages = {127–149}
}

@article{martinuzzi2025minimal,
  title = {Minimal deterministic echo state networks outperform random reservoirs in learning chaotic dynamics},
  volume = {35},
  ISSN = {1089-7682},
  url = {http://dx.doi.org/10.1063/5.0288751},
  DOI = {10.1063/5.0288751},
  number = {9},
  journal = {Chaos: An Interdisciplinary Journal of Nonlinear Science},
  publisher = {AIP Publishing},
  author = {Martinuzzi,  Francesco},
  year = {2025},
  month = sep 
}

@article{leutbecher2008ensemble,
  title = {Ensemble forecasting},
  volume = {227},
  ISSN = {0021-9991},
  url = {http://dx.doi.org/10.1016/j.jcp.2007.02.014},
  DOI = {10.1016/j.jcp.2007.02.014},
  number = {7},
  journal = {Journal of Computational Physics},
  publisher = {Elsevier BV},
  author = {Leutbecher,  M. and Palmer,  T.N.},
  year = {2008},
  month = mar,
  pages = {3515–3539}
}

@article{tebaldi2007use,
  title = {The use of the multi-model ensemble in probabilistic climate projections},
  volume = {365},
  ISSN = {1471-2962},
  url = {http://dx.doi.org/10.1098/rsta.2007.2076},
  DOI = {10.1098/rsta.2007.2076},
  number = {1857},
  journal = {Philosophical Transactions of the Royal Society A: Mathematical,  Physical and Engineering Sciences},
  publisher = {The Royal Society},
  author = {Tebaldi,  Claudia and Knutti,  Reto},
  year = {2007},
  month = jun,
  pages = {2053–2075}
}

@article{palmer2005representing,
  title = {Representing model uncertainty in weather and climate prediction},
  volume = {33},
  ISSN = {1545-4495},
  url = {http://dx.doi.org/10.1146/annurev.earth.33.092203.122552},
  DOI = {10.1146/annurev.earth.33.092203.122552},
  number = {1},
  journal = {Annual Review of Earth and Planetary Sciences},
  publisher = {Annual Reviews},
  author = {Palmer,  T.N. and Shutts,  G.J. and Hagedorn,  R. and Doblas-Reyes,  F.J. and Jung,  T. and Leutbecher,  M.},
  year = {2005},
  month = may,
  pages = {163–193}
}

@article{Jaeger_mme_mention_2007,
title = {Optimization and applications of echo state networks with leaky- integrator neurons},
journal = {Neural Networks},
volume = {20},
number = {3},
pages = {335-352},
year = {2007},
note = {Echo State Networks and Liquid State Machines},
issn = {0893-6080},
doi = {https://doi.org/10.1016/j.neunet.2007.04.016},
url = {https://www.sciencedirect.com/science/article/pii/S089360800700041X},
author = {Herbert Jaeger and Mantas Lukoševičius and Dan Popovici and Udo Siewert}
}

@article{Casanova_2023_hadron_application,
	author = {{Casanova, Maxime} and {Dalena, Barbara} and {Bonaventura, Luca} and {Giovannozzi, Massimo}},
	title = {Ensemble reservoir computing for dynamical systems: prediction of phase-space stable region for hadron storage rings},
	DOI= "10.1140/epjp/s13360-023-04167-y",
	url= "https://doi.org/10.1140/epjp/s13360-023-04167-y",
	journal = {Eur. Phys. J.  Plus},
	year = 2023,
	volume = 138,
	number = 6,
	pages = "559",
}

@inproceedings{lorenz1996predictability,
  title={Predictability: A problem partly solved},
  author={Lorenz, Edward N},
  booktitle={Proc. Seminar on predictability},
  volume={1},
  number={1},
  pages={1--18},
  year={1996},
  organization={Reading}
}

@article{Chiara_ref_lorenz96_params,
title = {Heterogeneity of the attractor of the Lorenz ’96 model: Lyapunov analysis, unstable periodic orbits, and shadowing properties},
journal = {Physica D: Nonlinear Phenomena},
volume = {457},
pages = {133970},
year = {2024},
issn = {0167-2789},
doi = {https://doi.org/10.1016/j.physd.2023.133970},
url = {https://www.sciencedirect.com/science/article/pii/S016727892300324X},
author = {Chiara Cecilia Maiocchi and Valerio Lucarini and Andrey Gritsun and Yuzuru Sato}
}

@article{MME_positive_2016,
title = {A review of multimodel superensemble forecasting for weather, seasonal
climate, and hurricanes},
journal = {Rev. Geophys.},
volumne = {54},
pages = {336-377},
year = {2016},
doi = {doi:10.1002/2015RG000513},
author = {T.N. Krishnamurti and V.Kumar and A.Simon and A.Bhardwaj and T.Ghosh and R.Ross}
}

@article{Weigel_risk,
    author = {Andreas P. Weigel and Reto Knutti and Mark A. Liniger and Christof Appenzeller},
    title = {Risks of Model Weighting in Multimodel Climate Projections},
    journal = {Journal of Climate},
    volumne = {23},
    pages = { 4175-4191},
    year = {2010},
    doi = {https://doi.org/10.1175/2010JCLI3594.1}
}

@article{Wei_comparative_study,
    author = {Wei, X. and Sun, X. and Sun, J. and Yin, J. and Sun, J. and Liu, C},
    title = {A Comparative Study of Multi-Model Ensemble Forecasting Accuracy between
Equal- and Variant-Weight Techniques},
    journal = {Atmosphere} ,
    year = {2022},
    volumne = {13(4)},
    pages = {526},
    doi = {https://doi.org/10.3390/atmos13040526}
}

@article{FTLE,
    author = {G. Boffetta and M. Cencini and M. Falcioni and A. Vulpiani},
	doi = {10.1016/S0370-1573(01)00025-4},
	issn = {0370-1573},
	journal = {Physics Reports},
    title = {Predictability: a way to characterize complexity},
    year = {2002},
    volumne = {356},
    pages = {367-474},
    doi = {https://doi.org/10.1016/S0370-1573(01)00025-4}
}

@article{transienttimes,
  title = {Transient times and cycle-rich topology in reservoir computing},
  author = {D. Estevez-Moya and M. Chai
  and E.A. Madrigal Solis and
  E. Estevez-Rams and H. Kantz},
  journal = {Chaos},
  volumne = {36},
  pages = {043104},
  year = {2026},
  doi = {https://doi.org/10.1063/5.0302415}
}

@article{Griffith_2019_distribution_errors,
    author = {Griffith, Aaron and Pomerance, Andrew and Gauthier, Daniel J.},
    title = {Forecasting chaotic systems with very low connectivity reservoir computers},
    journal = {Chaos: An Interdisciplinary Journal of Nonlinear Science},
    volume = {29},
    number = {12},
    pages = {123108},
    year = {2019},
    month = {12},
    issn = {1054-1500},
    doi = {10.1063/1.5120710},
    url = {https://doi.org/10.1063/1.5120710},
}

@article{GrassbergerProcacciaD2,
    author = {P. Grassberger, I. Procaccia},
    title = {Measuring the strangeness of strange attractors},
    journal = {Physica D},
    volumne = {9},
    pages = {189-208},
    year = {1983}
}

@article{Platt2022,
  title = {A systematic exploration of reservoir computing for forecasting complex spatiotemporal dynamics},
  volume = {153},
  ISSN = {0893-6080},
  url = {http://dx.doi.org/10.1016/j.neunet.2022.06.025},
  DOI = {10.1016/j.neunet.2022.06.025},
  journal = {Neural Networks},
  publisher = {Elsevier BV},
  author = {Platt,  Jason A. and Penny,  Stephen G. and Smith,  Timothy A. and Chen,  Tse-Chun and Abarbanel,  Henry D.I.},
  year = {2022},
  month = Sept,
  pages = {530–552}
}

@article{yildiz2012revisiting,
  title   = {Re-visiting the echo state property},
  author  = {Yildiz, Izzet B. and Jaeger, Herbert and Kiebel, Stefan J.},
  year    = {2012},
  journal = {Neural Networks},
  volume  = {35},
  pages   = {1--9},
  doi     = {10.1016/j.neunet.2012.07.005}
}

@article{manjunath2013echo,
  title   = {Echo state property linked to an input: Exploring a fundamental characteristic of recurrent neural networks},
  author  = {Manjunath, G. and Jaeger, H.},
  year    = {2013},
  journal = {Neural Computation},
  volume  = {25},
  number  = {3},
  pages   = {671--696},
  doi     = {10.1162/NECO_a_00411}
}

\end{document}